\documentclass[pra,preprint,showpacs,showkeys]{revtex4}
\usepackage{graphicx}
\usepackage{amsmath}
\usepackage{amssymb}
\usepackage{dsfont}

\newcommand{\ket}[1]{\left\vert#1\right\rangle}
\newcommand{\bra}[1]{\left\langle#1\right\vert}
\newcommand{\inprod}[2]{\left\langle#1\vert#2\right\rangle}
\newcommand{\abs}[1]{\left\vert#1\right\vert}

\begin{document}

\title{Deterministic Transformations of Multipartite Entangled States with Tensor Rank 2}
\author{S. Turgut}
\author{Y. G\"ul}
\author{N. K. Pak}
\affiliation{
Department of Physics, Middle East Technical University,\\
06531, ANKARA, TURKEY}

\begin{abstract}
Transformations involving only local operations assisted with
classical communication are investigated for multipartite
entangled pure states having tensor rank 2. All necessary and
sufficient conditions for the possibility of deterministically
converting truly multipartite, rank-2 states into each other are
given. Furthermore, a chain of local operations that successfully
achieves the transformation has been identified for all allowed
transformations. The identified chains have two nice features: (1)
each party needs to carry out at most one local operation and (2)
all of these local operations are also deterministic
transformations by themselves. Finally, it is found that there are
disjoint classes of states, all of which can be identified by a
single real parameter, which remain invariant under deterministic
transformations.
\end{abstract}



\maketitle

\section{Introduction}

Entanglement is a physical resource that enables one to carry out
classically impossible tasks such as teleportation\cite{Bennett_Telep}
and dense coding\cite{Bennett_Densecode}. As
such tasks require only special entangled states to be used in
their implementation, the transformation of entanglement has
become a major problem that has been widely studied.
Entanglement purification\cite{Bennett_Purif} is an example of such transformations
where mixed states, which are necessarily produced by noisy
quantum communication channels when the entangled particles
are distributed to distant parties, are converted into pure states. The main
problem in entanglement transformations is to understand the conditions and the necessary
protocols of the conversion process of a given state by local
quantum operations assisted with classical communication (LOCC) to
another desired state.

The transformations between pure bipartite-entangled states have been
understood best due to the existence of the Schmidt decomposition of such states.
The first important result on such transformations is the pioneering work of
Bennett \emph{et al.}\cite{Bennett_Conc} on the asymptotic transformations where
multiple copies of the same state are needed to be converted, which
establishes the entropy of entanglement as the sole currency of
conversion. The next major step was taken by Lo and Popescu\cite{lo_popescu}
who laid much of the groundwork for the
transformation of single copies of pure bipartite states.
Subsequently, Nielsen\cite{Nielsen_Maj} discovered the rules
of deterministic transformations where a simple connection between
the entanglement transformations and the mathematical theory of
majorization is established. Based on these developments, the conditions for
probabilistic transformations have also been determined\cite{Vidal_Prob,Plenio_Prob}.

In contrast to the bipartite case, Schmidt decomposition is not
available for multipartite entanglement between three or more
particles\cite{Peres_Schmidt}, and hence not much is known about
the transformations of such states. Some general results about the
transformations of multipartite pure states are given in Ref.
\onlinecite{Bennett_Multi}. Apart from this, all known
transformation rules are obtained for a restricted class of
states. For example, it has been shown that if the given and
desired states in question have a Schmidt decomposition, then the
transformation rules for bipartite states can be directly
applied\cite{Xin_Duan}. There are also works focusing on states
that lack a Schmidt decomposition. Namely, the probabilistic
distillation of the tripartite GHZ state\cite{GHZ_Distill} and
some aspects of the deterministic transformation between GHZ class
states\cite{Spedalieri} of three qubits have been studied. A
systematic treatment of transformations of this kind of state is
the subject of this article.

First, consider $p$ particles distributed to $p$ distant persons
(parties) where $p\geq3$ and let $\ket{\psi}$ be a state of these particles.
This state can be written as a sum of product states as
\begin{equation}
  \ket{\psi}= \sum_{i=1}^r \ket{\varphi_i^{(1)}\otimes  \varphi_i^{(2)}\otimes \cdots\otimes\varphi_i^{(p)}}
\end{equation}
where $\left\{\ket{\varphi_i^{(k)}} \right\}_{i=1}^r$ are vectors
in the state space of the $k$th particle (but these $r$ vectors
are not necessarily orthogonal to each other). The minimum possible
value of the number of terms $r$ in that expression is called the
\emph{tensor rank} of the state $\ket{\psi}$. The tensor rank is
sometimes also called the \emph{Schmidt rank}, and its base-2
logarithm gives the \emph{Schmidt measure} of the multipartite
states\cite{Duan_tensor,Eisert_Briegel}.

The main subject of this article is states with tensor rank 2,
which will be simply called as rank 2 states. As the matrix rank
of the reduced density matrices for each party is at most 2, these
states can always be considered as states of $p$ qubits. They can
be obtained by LOCC with non-zero probability from the generalized
GHZ state
\begin{equation}
  \ket{\mathrm{GHZ}} =  \frac{1}{\sqrt{2}}\left(\ket{0,0,\ldots,0}+\ket{1,1,\ldots,1}\right)\quad.
  \label{eq:GHZ}
\end{equation}
But note that they also contain states that cannot be used to
distill the GHZ state, i.e., states where some parties are
unentangled from the rest, even some bipartite entangled states
are also included in the set of rank 2 states.



The rank 2 states can be considered as the simplest type of entangled
states. Since any local operation on such states produces either a
rank 2 or a product state, the analysis of their transformations
should be somewhat simpler than the ones on other multipartite
states. Since they also lack a Schmidt decomposition, it will be
interesting to investigate their transformations. The purpose of
this article is twofold. First, it proposes a simple
parametrization of rank 2 states. This parametrization essentially
does the same job that the Schmidt decomposition does for the
analysis of bipartite states: it simplifies the identification of
states that can be converted to each other by local unitaries, an
essential task in the analysis of transformations. It also
simplifies the analysis and parametrization of local operations.
Second purpose of this article is to completely describe the
deterministic transformations between truly multipartite rank 2
states. It is hoped that, the results obtained in this article
will shed light on future studies on transformations between more
complicated states.

The organization of the article is as follows. In section
\ref{sec:param}, a parametrization of the rank 2 states is given
and the equivalence relation under the local unitaries is described.
After that, using the established parametrization of states,
local quantum operations that can be carried out by each
party are described and two possible parametrizations of these operations are proposed.
In section \ref{sec:deter}, the necessary and
sufficient conditions for the possibility of deterministic
transformations between two given multipartite states are
obtained. Finally, a brief conclusion is given in \ref{sec:conc}.

\section{The Description of the States and the Local Operations}
\label{sec:param}

%

\subsection{The parametrization of states with ranks 1 and 2}

By definition, any rank-2 state $\ket{\psi}$ can be expressed in
the form
\begin{equation}
  \ket{\psi}=\frac{1}{\sqrt{N}}
  \left( \ket{\alpha_1\otimes\alpha_2\otimes\cdots\otimes\alpha_p}
         + z \ket{\beta_1\otimes\beta_2\otimes\cdots\otimes\beta_p}
  \right)
\label{eq:eq_psi_defined}
\end{equation}
where $\ket{\alpha_k}$ and $\ket{\beta_k}$ are \emph{normalized}
states in the Hilbert space $\mathcal{H}_k$ of the particle
possessed by the party-$k$, where their relative phases are
adjusted suitably such that they have a real, non-negative inner
product $c_k=\inprod{\alpha_k}{\beta_k}$ (i.e., $c_k\geq0$), and
$z$ is a complex number. Here $N$ is a normalization factor. The
overall phase of the state $\ket{\psi}$ can be eliminated by
absorbing it into the overall phase of a pair
$\{\ket{\alpha_k},\ket{\beta_k}\}$ for one of the parties.

Apart from the vectors $\ket{\alpha_k}$ and $\ket{\beta_k}$, this
state depends on one complex parameter $z$, and $p$ real
parameters $c_1,\ldots,c_p$, which will simply be called as the
\emph{cosines}, belonging to the closed interval $[0,1]$. The
collection of these parameters will be denoted by
$\lambda=(z;c_1,c_2,\ldots,c_p)$. The $(p+1)$-tuple $\lambda$ will
be considered as a point in a space $\Lambda$ which is essentially
$\mathbb{C}\times[0,1]^p$. However, there are a few adjustments to
be made before defining the space $\Lambda$ precisely. First, if
$\ket{\psi}$ is a rank-2 state, then the complex number $z$ has to
be non-zero. However, treating the product states (i.e., rank-1
states) by the same parametrization has some advantages. For this
reason, $z=0$ values are also included as possible values of this
parameter. Moreover, the value $z=\infty$ should also be included
as a possible value for this parameter, where $\ket{\psi}$ is
again a product state. In other words, the parameter $z$ can be
chosen from the extended complex numbers
$\mathbb{C}^\prime=\mathbb{C}\cup\{\infty\}$. Apart from this,
note that the point $(z=-1;c_1=1,\ldots,c_p=1)$ cannot possibly be
identified with a state. As a result, this point is excluded from
$\Lambda$. Hence, the parameter space $\Lambda$ is defined as
$\Lambda=\mathbb{C}^\prime\times[0,1]^p \setminus
\{(-1;1,1,\ldots,1)\}$. Consequently, any rank-2 or rank-1 state
can be expressed by using a point $\lambda$ in so-defined space
$\Lambda$.

It is possible to distinguish three types of states that can be
represented as a point in $\Lambda$. The following list describes
these types and gives the necessary rules for understanding the
type of the state a point $\lambda=(z;c_1,\ldots,c_p)$ represents.
\begin{itemize}

\item[(1)] The product states, which are included into $\Lambda$
for completeness due to the fact that some local operations
produce them. The point $\lambda$ corresponds to a product state
if and only if either (i) $z=0$, or (ii) $z=\infty$, or (iii) the
cosines of at least $p-1$ parties are $1$.

\item[(2)] Bipartite entangled states.  A point $\lambda$ is a
bipartite entangled state if and only if $z\neq0,\infty$ and
\emph{exactly} $p-2$ of the cosines are $1$ and the remaining two
cosines are strictly less than $1$. If $\lambda$ is a bipartite
state between parties $k_1$ and $k_2$, then we should have
$c_{k_1},c_{k_2}<1$. For all of these states, the cosines can be
chosen in a multitude of different ways, proving that the current
parametrization is rather inconvenient for these types of states.

\item[(3)] The rest of the states, i.e., those that are
neither product nor bipartite entangled, will be called
\emph{truly multipartite} states. The point $\lambda$ corresponds
to a truly multipartite state if and only if $z\neq0,\infty$ and
at least three of the cosines are strictly less than $1$.
\end{itemize}

Note that the concurrence\cite{Wootters_Concur} can be used as a
measure of the entanglement of a given party $k$ with all the
other parties. When the state $\ket{\psi}$ in
Eq.~(\ref{eq:eq_psi_defined}) is considered as an entangled state
between the party-$k$ and the rest of the parties, the related
concurrence can be computed as,
\begin{equation}
  \mathcal{C}_k = 2\sqrt{\det\rho^{(k)}}
\end{equation}
where $\rho^{(k)}$ is the reduced density matrix for the party
$k$, which is defined by
$\rho^{(k)}=\mathrm{tr}_{1,2,\ldots,k-1,k+1,\ldots,p}\ket{\psi}\bra{\psi}$.
It is straightforward to compute these concurrences for the state
in Eq.~(\ref{eq:eq_psi_defined}) as
\begin{equation}
   \mathcal{C}_k = \frac{2\abs{z}\sqrt{1-c_k^2}\sqrt{1-(c_1\cdots c_{k-1}c_{k+1}\cdots c_p)^2}}{N}\quad.
\end{equation}
Note that for $z\neq0,\infty$, the concurrence $\mathcal{C}_k$ can
be non-zero (and hence the party-$k$ is entangled with the others)
if and only if $c_k<1$ and there is another cosine which is less
than $1$. Using this, all of the rules listed above can be
justified easily.

\subsection{Local Unitary Equivalence}

Next, a description of local unitary (LU) equivalence between
states that are expressed by using the parametrization given above
must be provided. Two states $\ket{\psi}$ and $\ket{\phi}$ are
called LU-equivalent, if there are local unitary operators $V_k$
on each local Hilbert space $\mathcal{H}_k$ such that
$\ket{\psi}=(V_1\otimes\cdots\otimes V_p)\ket{\phi}$. Obviously,
LU-equivalent states can be converted into each other by LOCC,
with necessary local quantum operations being the indicated
unitaries. The opposite is also true. If two states $\ket{\psi}$
and $\ket{\phi}$ are LOCC-convertible into each other, then
$\ket{\psi}$ and $\ket{\phi}$ are LU equivalent\cite{Bennett_Multi}.


It is obvious that any two states described by the same parameter
$\lambda=(z;c_1,\ldots,c_p)$ are LU-equivalent. Such states differ
only in the pairs of states $\{\ket{\alpha_k},\ket{\beta_k}\}$
which have a common inner product $c_k$, and therefore it is
possible find a unitary operator that converts such pairs into
similar pairs. Hence, the point $\lambda=(z;c_1,\ldots,c_p)$
denotes a collection states which are all LU-equivalent to the
following representative state
\begin{equation}
  \ket{\Phi(\lambda)} = \frac{1}{\sqrt{N(\lambda)}}
   \left(\ket{0\otimes0\otimes\cdots\otimes0}+z\ket{w_{c_1}\otimes w_{c_2}\otimes\cdots\otimes w_{c_p}}\right)
\end{equation}
where $\ket{w_{c_k}}=c_k\ket{0}+\sqrt{1-c_k^2}\ket{1}$ and
\begin{equation}
  N(\lambda)=1+\abs{z}^2+c_1c_2\cdots c_p(z+z^*)\quad.
\end{equation}

Apart from this, it is possible to express the \emph{same} state
by using two different points of $\Lambda$, say $\lambda$ and
$\lambda^\prime$. Expressed in a different but equivalent way: there might be
different points $\lambda$ and $\lambda^\prime$ of $\Lambda$ such
that the representative states $\ket{\Phi(\lambda)}$ and
$\ket{\Phi(\lambda^\prime)}$ are LU-equivalent. The points
$\lambda$ and $\lambda^\prime$ will be called (LU) equivalent if
this happens and that relation will be denoted as $\lambda \sim
\lambda^\prime$.

For any given point $\lambda=(z;c_1,\ldots,c_p)$, let us define
its \emph{conjugate} point by
$\hat{\lambda}=(1/z;c_1,\ldots,c_p)$, i.e., point obtained by
inverting the $z$-parameter only. It can be seen easily that
$\lambda \sim \hat{\lambda}$. It turns out that, if none of the
cosines of $\lambda$ vanish, then the equivalence class of
$\lambda$ is formed by the pair of points
$\{\lambda,\hat{\lambda}\}$. Precise criteria for deciding whether
two given points of $\Lambda$ are equivalent are given below.

Let $\lambda=(z;c_1,c_2,\ldots,c_p)$ and
$\lambda^\prime=(z^\prime;c_1^\prime,c_2^\prime,\ldots,c_p^\prime)$.
The rules of equivalence depend on the types of the states as
follows.
\begin{itemize}
\item[(1)] If $\lambda$ is product state, then $\lambda \sim
\lambda^\prime$ if and only if $\lambda^\prime$ is a product
state.

\item[(2)] If $\lambda$ is a bipartite entangled state between
party-$k_1$ and party-$k_2$, then $\lambda \sim \lambda^\prime$ if
and only if $\lambda^\prime$ is also a bipartite state between the
same parties and they have the same concurrences, in other words
\begin{equation}
 \frac{\abs{z}\sqrt{1-c_{k_1}^2}\sqrt{1-c_{k_2}^2}}{N(\lambda)}=
 \frac{\abs{z^\prime}\sqrt{1-c_{k_1}^{\prime2}}\sqrt{1-c_{k_2}^{\prime2}}}{N(\lambda^\prime)}    ~.
\label{eq:concurrence_eq_bipartite}
\end{equation}

\item[(3)] If $\lambda$ is truly multipartite then
$\lambda\sim\lambda^\prime$ if and only if (a) the corresponding
cosines are identical
(i.e., $c_k=c_k^\prime$ for all $k$) and (b) the following condition holds for $z$
depending on whether there is a vanishing cosine or not:
\begin{itemize}
 \item[(i)] when no cosines vanish: either $z^\prime=z$ or $z^\prime=1/z$,
 \item[(ii)] when there is a vanishing cosine: either $\vert
z^\prime\vert=\vert z\vert$ or $\vert z^\prime\vert=1/\vert
z\vert$.
\end{itemize}

\end{itemize}

The statements for the product and bipartite states are
straightforward. The last rule follows simply from the following
theorem.\\
\emph{Theorem 1. Let $\ket{\psi}$ be a state of $p$ particles ($p\geq2$)
which is expressed as
\begin{equation}
  \ket{\psi}= \sum_{i=1}^r \ket{\varphi_i^{(1)}\otimes  \varphi_i^{(2)}\otimes \cdots\otimes\varphi_i^{(p)}}
\end{equation}
where, for each $k$,
$F^{(k)}=\left\{\ket{\varphi_1^{(k)}},\ket{\varphi_2^{(k)}},\ldots,\ket{\varphi_r^{(k)}}\right\}$
is a set of $r$ \emph{non-zero}, possibly unnormalized
vectors from the Hilbert space $\mathcal{H}_k$. Let $m$ denote the
number of parties $k$ for which the set $F^{(k)}$ is linearly
independent. Then the following statements hold.
\begin{itemize}
\item[(a)] If $m\geq1$, then the set of $r$ vectors
$G=\left\{\ket{\varphi_i^{(1)}\otimes\varphi_i^{(2)}\otimes\cdots\otimes\varphi_i^{(p)}}\right\}_{i=1}^r$
is also linearly independent.
\item[(b)] If $F^{(\ell)}$ is linearly independent, then for all
$k \neq\ell$, $\mathrm{supp}\,\rho^{(k)}=\mathrm{span}\,F^{(k)}$.
\item[(c)] If $m\geq2$, then $\ket{\psi}$ has tensor rank $r$.
\item[(d)] If $m\geq3$,  then the expression of $\ket{\psi}$ as a
superposition of $r$ product states is unique. In other words, if
$\ket{\beta_i^{(k)}}$ are vectors such that
\begin{equation}
  \ket{\psi} = \sum_{i=1}^r \ket{\beta_i^{(1)}\otimes  \beta_i^{(2)}\otimes \cdots\otimes \beta_i^{(p)}}
\end{equation}
then there is a permutation $Q$ of $r$ objects such that for any
$i=1,\ldots,r$ we have
\begin{equation}
  \ket{\beta_i^{(1)}\otimes  \beta_i^{(2)}\otimes \cdots\otimes \beta_i^{(p)}}=\ket{\varphi_{Qi}^{(1)}\otimes  \varphi_{Qi}^{(2)}\otimes \cdots\otimes \varphi_{Qi}^{(p)}}\quad.
\end{equation}
\end{itemize}
}

\textit{Proof:} To simplify the proof, it can be assumed that the
parties are relabelled so that the first $m$ sets of vectors,
i.e., $F^{(1)},\ldots,F^{(m)}$, are linearly independent. This
assumption is employed in all of the cases treated below.

For (a), suppose that there are numbers $a_1,a_2,\ldots,a_r$ such
that
\begin{equation}
  \sum_{i=1}^r a_i \ket{\varphi_i^{(1)}\otimes  \varphi_i^{(2)}\otimes \cdots  \otimes\varphi_i^{(p)}} =0\quad.
\end{equation}
Let $\ket{\Theta}=\ket{\Theta}_{2\ldots p}$ be an arbitrary vector
in the Hilbert space of all parties except the $1$st. The inner
product with this state gives
\begin{equation}
  \sum_{i=1}^r \ket{\varphi_i^{(1)}}\left(a_i\inprod{\Theta}{\varphi_i^{(2)}\otimes \cdots  \otimes\varphi_i^{(p)}}\right) =0\quad.
\end{equation}
Due to linear independence of $F^{(1)}$, we have
$a_i\inprod{\Theta}{\varphi_i^{(2)}\otimes \cdots
\otimes\varphi_i^{(p)}}=0$  for all $i$. As $\ket{\Theta}$ is
arbitrary and each $\ket{\varphi^{(k)}_i}$ is non-zero, it
necessarily follows that $a_i=0$. This shows that the set of
vectors $G$ is linearly independent.

For (b), suppose that $\ell=1$ and $k=2$ without loss of
generality. The reduced density matrix for the second party is
\begin{equation}
  \rho^{(2)}=\sum_{i,j=1}^r S_{ji} \ket{\varphi_i^{(2)}}\bra{\varphi_j^{(2)}}\quad,
\end{equation}
where $S$ is the overlap matrix given by
$S_{ji}=\inprod{\chi_j}{\chi_i}$ and
$\ket{\chi_i}=\ket{\varphi_i^{(1)}\otimes  \varphi_i^{(3)}\otimes
\cdots \otimes\varphi_i^{(p)}}$. Using the result in (a), it can
be seen that the set of $r$-vectors
$\left\{\ket{\chi_i}\right\}_{i=1}^r$ is also linearly
independent. Therefore, the overlap matrix $S$ is strictly
positive definite.

From the expression of $\rho^{(2)}$, it is obvious that the
support of $\rho^{(2)}$ is included in $\mathrm{span}\,F^{(2)}$.
To show that these two subspaces are identical, let us assume the
contrary. Let $\ket{\varphi^\prime}$ be a non-zero vector in
$\mathrm{span}\,F^{(2)}$ but orthogonal to the support of
$\rho^{(2)}$. Then, at least one of
$b_i=\inprod{\varphi_i^{(2)}}{\varphi^\prime}$ is non-zero and
therefore
$\bra{\varphi^\prime}\rho^{(2)}\ket{\varphi^\prime}=\sum_{ij}
S_{ji}b_j^* b_i>0$, which is a contradiction. This then shows that
the two subspaces are identical, i.e.,
$\mathrm{supp}\,\rho^{(2)}=\mathrm{span}\,F^{(2)}$.

For (c), note that both $F^{(1)}$ and $F^{(2)}$ are linearly
independent and hence by (b),
$\mathrm{supp}\,\rho^{(k)}=\mathrm{span}\,F^{(k)}$ for all parties
$k$. In particular, $\rho^{(1)}$ has matrix rank $r$. This shows
that $\ket{\psi}$ cannot be written as a sum of product states
with less than $r$ terms. Hence, the tensor rank of $\ket{\psi}$
is $r$.

Finally, consider the statement in part (d). Let
$B^{(k)}=\left\{\ket{\beta_1^{(k)}},\ket{\beta_2^{(k)}},\ldots,\ket{\beta_r^{(k)}}\right\}$.
Since the tensor rank of $\ket{\psi}$ is $r$, all of the vectors
$\ket{\beta_i^{(k)}}$ are non-zero. Moreover, since the matrix
rank of $\rho^{(k)}$ is $r$ for $k=1,2,3$, the sets $B^{(1)}$,
$B^{(2)}$ and $B^{(3)}$ are linearly independent. In short, the
prerequisite conditions for the theorem and part (d) are satisfied
for the new vectors $\ket{\beta_i^{(k)}}$ as well. Hence we have
$\mathrm{span}\,B^{(k)}=\mathrm{span}\,F^{(k)}$.

Since $\mathrm{span}\,B^{(1)}=\mathrm{span}\,F^{(1)}$ and
$B^{(1)}$ is also linearly independent, there is an $r\times r$
invertible matrix $Z$ such that
\begin{equation}
  \ket{\beta_i^{(1)}} = \sum_{j=1}^r Z_{ji}  \ket{\varphi_j^{(1)}}\quad.
\end{equation}
Inserting this into the expansions of $\ket{\psi}$ we get
\begin{equation}
  \sum_j \ket{\varphi_j^{(1)}\otimes\varphi_j^{(2)}\otimes\cdots\otimes\varphi_j^{(p)}}
  =\sum_{ij}Z_{ji}\ket{\varphi_j^{(1)}\otimes\beta_i^{(2)}\otimes\cdots\otimes\beta_i^{(p)}}\quad.
\end{equation}
Using the linear independence of $F^{(1)}$, we get
\begin{equation}
 \ket{\varphi_j^{(2)}\otimes\cdots\otimes\varphi_j^{(p)}}
  = \sum_{i=1}^r  Z_{ji}\ket{\beta_i^{(2)}\otimes\cdots\otimes\beta_i^{(p)}}
\end{equation}
which must hold true for all $r$. In here, a rank-1 state (product
state) is expanded as a sum of $r$ product states. Note that
$B^{(2)}$ and $B^{(3)}$ are both linearly independent and
therefore part (c) of the current theorem can be applied to this
expression. It then directly follows that only one number in the
sequence $Z_{j1},Z_{j2},\ldots,Z_{jr}$ can be non-zero (otherwise
we get a contradiction for the tensor rank of the state on the
left-hand side). As $Z$ is a square matrix, each row and each
column contains only one non-zero entry.

Let $Q$ be the permutation that gives the index of the non-zero
entry for a given column. In other words, $Z_{ji}\neq0$ only for
$j=Q_i$. Then, we have
\begin{eqnarray}
  \ket{\beta_i^{(1)}} &=&  Z_{Q_ii}  \ket{\varphi_{Q_i}^{(1)}}  \quad,\\
  \ket{\varphi_{Q_i}^{(2)}\otimes\cdots\otimes\varphi_{Q_i}^{(p)}}
      &=& Z_{Q_ii}\ket{\beta_i^{(2)}\otimes\cdots\otimes\beta_i^{(p)}}\quad,\\
  \Longrightarrow\ket{\beta_i^{(1)}\otimes\beta_i^{(2)}\otimes \cdots\otimes \beta_i^{(p)}}
      &=& \ket{\varphi_{Qi}^{(1)}\otimes  \varphi_{Qi}^{(2)}\otimes \cdots\otimes \varphi_{Qi}^{(p)}}\quad,
\end{eqnarray}
which is what is aimed to be proved.$\Box$

At this point, let us briefly investigate the implication of the
theorem for the rank-2 states. Let $\ket{\psi}$ be the state
defined by Eq.~(\ref{eq:eq_psi_defined}) and suppose that
$z\neq0,\infty$. In this case, the conditions in the theorem are
satisfied with $r=2$. Here, the statement that the set
$F^{(k)}=\{\ket{\alpha_k},\ket{\beta_k}\}$ is linearly independent
is equivalent to $c_k<1$. Hence, part (c) is the case for truly
multipartite states. As a result, the expansion of $\ket{\psi}$ in
Eq.~(\ref{eq:eq_psi_defined}) is unique. The most one can do in
here is to exchange the places of the two terms, which essentially
changes $\lambda$ to $\hat{\lambda}$. Consequently, the cosines of
individual parties do not change. The rest of the rule follows
trivially.

\subsection{Local Operations by a Single Party}

A local quantum operation (measurement) applied by a single party-$k$ can be
described by the general measurement formalism, i.e., there is a
set of possible outcomes and for each outcome $\ell$, there is an
associated measurement operator $M_\ell$ on $\mathcal{H}_k$, all of
which satisfy the probability-sum condition
\begin{equation}
  \sum_{\ell} M_\ell^\dagger M_\ell=\mathds{1}_k\quad.
  \label{eq:ProbabilitySumRule}
\end{equation}
If the state before the operation is  $\ket{\psi}$, then the
outcome $\ell$ occurs with probability
$p_\ell=\bra{\psi}(M_\ell^\dagger
M_\ell)\otimes\mathds{1}_k^\prime\ket{\psi}$ and the state
collapses to
$\ket{\psi^{(\ell)}}\propto(M_\ell\otimes\mathds{1}_k^\prime)\ket{\psi}$
(up to normalization), where $\mathds{1}_k^\prime$ denotes the
identity operator acting on all parties except the $k$th one. The
main purpose of this section is to provide two alternative
descriptions of the general measurement formalism that are more
suitable to work with when using the $\Lambda$ space for state
parametrization.

\subsubsection{First parametrization of local operations}
Suppose that the initial state $\ket{\psi}$ is given as in
Eq.~(\ref{eq:eq_psi_defined}) and consider a local operation
described by $M_\ell$. Define four real parameters for
each outcome $\ell$ as follows
\begin{eqnarray}
 A_\ell &=& \Vert M_\ell\ket{\alpha_k} \Vert \quad, \\
 B_\ell &=& \Vert M_\ell\ket{\beta_k} \Vert \quad, \\
 C_\ell e^{i\gamma_\ell} &=& \frac{1}{A_\ell B_\ell}\bra{\alpha_k} M_\ell^\dagger M_\ell \ket{\beta_k}\quad,
\end{eqnarray}
where the norm is defined as $\Vert \ket{\phi}\Vert=\sqrt{\inprod{\phi}{\phi}}$.
Here $C_\ell$ is taken to be a non-negative real number and
the phase $\gamma_\ell$ is defined accordingly. It follows that
$C_\ell\leq1$ by Schwarz inequality. The parameters $A_\ell$ and
$B_\ell$ are also necessarily non-negative. If the state before
the operation corresponds to the point
$\lambda=(z;c_1,\ldots,c_p)$, then the collapsed state for the outcome
$\ell$ corresponds to the point
$\lambda^{(\ell)}=(z^{(\ell)};c_1,\ldots,c_{k-1},C_\ell,c_{k+1},\ldots
c_p)$ where
\begin{equation}
  z^{(\ell)} = z \frac{B_\ell e^{i\gamma_\ell}}{A_\ell} \quad, \label{eq:FinalState_z}
\end{equation}
and the probability of that outcome is given by
\begin{eqnarray}
  p_\ell &=&  A_\ell^2\frac{N(\lambda^{(\ell)})}{N(\lambda)} \label{eq:FinalState_prob}\\
         &=& \frac{A_\ell^2+\abs{z}^2B_\ell^2
             + A_\ell B_\ell c_1\cdots c_{k-1}C_\ell c_{k+1}\cdots c_p
                   (ze^{i\gamma_\ell}+z^* e^{-i\gamma_\ell})}{N(\lambda)}\quad.
\end{eqnarray}

Hence, when describing the effect of a local operation by a single
party, the values of four real parameters for each outcome are
needed: $A_\ell$, $B_\ell$, $C_\ell$ and $\gamma_\ell$. Obviously,
possible values of these parameters are restricted by the
probability-sum condition (\ref{eq:ProbabilitySumRule}) and the
Schwarz inequality, which read
\begin{eqnarray}
  \sum_{\ell=1}^n A_\ell^2 = \sum_{\ell=1}^n B_\ell^2 &=&  1\quad, \label{eq:Par1}\\
  \sum_{\ell=1}^n A_\ell B_\ell C_\ell e^{i\gamma_\ell} &=& c_k\quad, \label{eq:Par2} \\
  C_\ell & \leq & 1\quad, \label{eq:Par3}
\end{eqnarray}
when party-$k$ is carrying out the operation.

It appears that these are the only restrictions on these
parameters. In other words, if a set of non-negative numbers
$A_\ell$, $B_\ell$, $C_\ell$ and angles $\gamma_\ell$ satisfy
Eqs.~(\ref{eq:Par1}-\ref{eq:Par3}), then it is possible to
construct measurement operators $M_\ell$ on the space
$\mathcal{H}_k$ which would produce the same set of parameters.
For showing this, consider only the case where $c_k<1$ (otherwise,
party $k$ is unentangled with the remaining parties and what she
does has no effect on the state). Let
$\{\ket{\alpha_k^\perp},\ket{\beta_k^\perp}\}$ be the dual basis
in $\mathcal{H}_k$, which satisfy
\begin{eqnarray}
\inprod{\alpha_k^\perp}{\alpha_k}=\inprod{\beta_k^\perp}{\beta_k} &=& 1\quad,\\
\inprod{\alpha_k^\perp}{\beta_k}=\inprod{\beta_k^\perp}{\alpha_k} &=& 0\quad.
\end{eqnarray}
The associated vectors are simply given by
\begin{eqnarray}
  \ket{\alpha_k^\perp} &=& \frac{1}{1-c_k^2}\left(\ket{\alpha_k}-c_k\ket{\beta_k}\right)\quad,\\
  \ket{\beta_k^\perp} &=& \frac{1}{1-c_k^2}\left(-c_k\ket{\alpha_k}+\ket{\beta_k}\right)\quad.
\end{eqnarray}
Next, define a new set of operators $P_\ell$ on $\mathcal{H}_k$ as
\begin{eqnarray}
  P_\ell &=& A_\ell^2\ket{\alpha_k^\perp}\bra{\alpha_k^\perp}+B_\ell^2 \ket{\beta_k^\perp}\bra{\beta_k^\perp}+ \nonumber\\
    & & \left(A_\ell B_\ell C_\ell e^{i\gamma_\ell} \ket{\alpha_k^\perp}\bra{\beta_k^\perp}+h.c.\right)~~.
\end{eqnarray}
It is straightforward to show that $P_\ell$ is positive
semidefinite (where the inequality $C_\ell\leq1$ is employed) and
$\sum_\ell P_\ell = \mathds{1}_k$ (where the remaining
restrictions, (\ref{eq:Par1}) and (\ref{eq:Par2}), are employed).
In short, the set of operators $\{P_\ell\}$ forms a
positive-operator valued measure (POVM). The measurement operators can be simply defined
as $M_\ell=\sqrt{P_\ell}$. It is then easy to check that the same
parameters are produced by these measurement operators. This completes
the proof that one only needs to satisfy the conditions (\ref{eq:Par1}-\ref{eq:Par3})
when employing the parametrization of local operations by party-$k$.

There are a number of remarks that should be made about the
parametrization of local measurements described above. First,
notice that this parametrization \emph{depends on the initial
parameter point $\lambda$} through the appearance of the cosine
$c_k$ in (\ref{eq:Par2}). Second, if an alternative, LU-equivalent
point is used for the initial point, then the parametrization of
the operation changes accordingly, even though it is the same
operation on the same state. For example, if $\hat{\lambda}$ is
used instead of $\lambda$, then the parameters of the operation
change as $A_\ell\leftrightarrow B_\ell$,
$\gamma_\ell\rightarrow-\gamma_\ell$ and $C_\ell$ remains same so
that $\lambda^{(\ell)}\rightarrow \widehat{\lambda^{(\ell)}}$.
Third, local operations do not change the cosines of the other
parties, i.e., if party-$k$ is carrying out the operation, then
$c_{k^\prime}$ remains the same for all $k^\prime\neq k$. The only
changes are in the cosine of party-$k$ (i.e., $c_k$ becomes
$C_\ell$ now) and the parameter $z$.

Fourth, any two different outcomes that produce the same final
point (e.g., outcomes $\ell_1\neq\ell_2$ with
$\lambda^{(\ell_1)}=\lambda^{(\ell_2)}$) can be combined to a
single outcome by choosing a new set of parameters. Hence, when
analyzing local operations, it can be supposed without loss of
generality that different outcomes correspond to different final
points of $\Lambda$. However, different outcomes corresponding to
different but LU-equivalent points of $\Lambda$ (e.g., outcomes
$\ell_1\neq\ell_2$ with $\lambda^{(\ell_1)}\neq\lambda^{(\ell_2)}$
but $\lambda^{(\ell_1)}\sim\lambda^{(\ell_2)}$) cannot in general
be combined to a single outcome by a simple redefinition of local
operation parameters.

Finally, the special outcomes where either $A_\ell=0$ or
$B_\ell=0$ produces a product state where $z^{(\ell)}$ is either
$0$ or $\infty$. The outcomes for which $A_\ell=B_\ell=0$ can be
simply discarded from consideration because they will always have
zero probability of occurrence.

A nice application of the first parametrization is the following
theorem which essentially expresses the idea that any non-unitary
local operation produces an outcome which is closer to the product
states.\\
\emph{Theorem 2. Consider a local operation by party-$k$ on a
state corresponding to point $\lambda=(z;c_1,\ldots,c_p)$. Then,
\begin{itemize}
 \item[(a)] There is at least one outcome $\ell$ for which $C_\ell\geq c_k$.
 \item[(b)] If $c_k\neq0$ and the local operation is not a set of random unitary transformations,
   then there is at least one outcome $\ell$ for which $C_\ell > c_k$.
 \item[(c)] If $\abs{z}\geq1$, there is at least one outcome $m$ for which $\abs{z^{(m)}}\geq \abs{z}$.
\end{itemize}
}

\textit{Proof:} The statement in (a) holds trivially for the
special case $c_k=0$; it also holds when the local operation is a
unitary transformation or a set of random unitary transformations
which never change the state parameters $\lambda$. Consequently,
proving (b) also proves (a). To prove (b), assume the contrary,
i.e., suppose that $c_k>0$ and the final cosines do not exceed
$c_k$ for all outcomes ($C_\ell\leq c_k$ for all $\ell$). Then,
\begin{eqnarray}
 c_k  &=& \sum_\ell A_\ell B_\ell C_\ell e^{i\gamma_\ell}\leq \sum_\ell A_\ell B_\ell C_\ell \\
    &\leq& c_k \sum_\ell A_\ell B_\ell \leq c_k\sqrt{(\sum_\ell A_\ell^2)(\sum_\ell B_\ell^2)}  = c_k
\end{eqnarray}
and, as a result, all inequalities must be equalities. Namely, we
should have $\gamma_\ell=0$ when $A_\ell B_\ell C_\ell\neq0$,
$C_\ell=c_k$ when $A_\ell B_\ell\neq0$ and $A_\ell=B_\ell$ from
the Schwarz inequality. The last relation rules out the
product-state producing outcomes (which are the cases where either
$A_\ell=0$ or $B_\ell=0$, but not both). Hence we have
$C_\ell=c_k$, $\gamma_\ell=0$ and $z^{(\ell)}=z$ for all $\ell$.
This means that all outcomes are identical and the state has not
changed. In other words, $k$th party has made a local unitary
transformation for all outcomes $\ell$, which is contrary to the
assumption.

For (c), again assume the contrary and suppose
$\abs{z^{(\ell)}} < \abs{z}$ for all outcomes $\ell$. This
implies that $B_\ell < A_\ell$. However,
\begin{equation}
  1=\sum_\ell B_\ell^2 < \sum_\ell A_\ell^2 =1~~,
\end{equation}
which is a contradiction. Therefore the statement in (c) holds.$\Box$


\subsubsection{Second parametrization of local operations}
For the local operation described above, the restrictions
(\ref{eq:Par1},\ref{eq:Par2}) imply the following identities
\begin{eqnarray}
  \sum_\ell p_\ell  \frac{1}{N(\lambda^{(\ell)})} &=&  \frac{1}{N(\lambda)}\quad, \label{eq:Rel1}\\
  \sum_\ell p_\ell  \frac{\abs{z^{(\ell)}}^2}{N(\lambda^{(\ell)})} &=&  \frac{\abs{z}^2}{N(\lambda)}\quad, \label{eq:Rel2} \\
  \sum_\ell p_\ell  \frac{z^{(\ell)}C_\ell}{N(\lambda^{(\ell)})} &=& \frac{z c_k}{N(\lambda)}\quad.\label{eq:Rel3}
\end{eqnarray}
An important feature of these relations is that they are expressed
entirely in terms of two real parameters (the probability $p_\ell$
and the final cosine $C_\ell$) and one complex parameter
($z^{(\ell)}$), which are all one needs for describing the effect
of the local operation. More importantly, these relations form a
basis for an alternative parametrization of the local operation by
party-$k$. In other words, if a set of outcomes are given and for
each outcome $\ell$, two real numbers, $p_\ell$ and $C_\ell$, and
one complex number, $z^{(\ell)}$, are given such that they
satisfy: (i) $\{p_\ell\}$ are probabilities, (ii) $0\leq
C_\ell\leq 1$ and (iii) the relations (\ref{eq:Rel1}) and
(\ref{eq:Rel3}) are satisfied, then it is possible to construct a
local operation for party-$k$ such that the final point
$\lambda^{(\ell)}=(z^{(\ell)};c_1,\ldots,c_{k-1},C_\ell,c_{k+1},\ldots,c_p)$
is obtained with probability $p_\ell$. To prove this, we simply
define
\begin{eqnarray}
  A_\ell &=& \sqrt{\frac{p_\ell N(\lambda)}{N(\lambda^{(\ell)})}}  \\
  B_\ell &=& \frac{\abs{z^{(\ell)}}}{\abs{z}}\sqrt{\frac{p_\ell N(\lambda)}{N(\lambda^{(\ell)})}} \\
  \gamma_\ell &=& \mathrm{arg}\,\left(z^{(\ell)}/z\right)
\end{eqnarray}
and check that Eq.~(\ref{eq:Par1}-\ref{eq:Par3}) are satisfied.
(The special cases $z^{(\ell)}=0,\infty$ or $z=0,\infty$ can be
handled by a limiting procedure without any problem.)

The second parametrization shares many features with the first.
For example, it depends on the initial point $\lambda$, different
outcomes corresponding to same point in $\Lambda$ can be combined
into a single outcome, etc. However, the second parametrization
is more convenient, and thus more useful because of the direct
appearance of the parameters of the final states in the conditions
(\ref{eq:Rel1}-\ref{eq:Rel3}).

\section{Deterministic Transformations of States by Many Parties}
\label{sec:deter}

Consider the general LOCC transformation by successive local
operations carried out by many parties on rank 2 states. Using
classical communication, the parties can coordinate their actions
depending on the outcomes of previous measurements. For the case
of deterministic transformations, it is required that all of the
possible final states are LU equivalent to each other. In this
section, deterministic transformation of an initial state
$\lambda$ to a final state $\lambda^\prime$ is considered. The
main problem is the determination of the necessary and sufficient
conditions for the possibility of such a transformation and the
design of a chain of local measurements when the transformation is
possible.

The special case where $\lambda$ corresponds to a bipartite state
falls into the scope of Nielsen's theorem\cite{Nielsen_Maj} and
is not needed to be considered in here. Moreover, the special case
where the final state $\lambda^\prime$ is a bipartite state
appears to be a complicated problem due to the presence of an
enormous number of points in the LU-equivalence class of all
bipartite states in $\Lambda$. Hence, in the rest of this section,
it will be assumed that both of the initial and the final states
are truly multipartite.

Because of the LU-equivalences $\lambda\sim\hat{\lambda}$ and
$\lambda^\prime\sim\hat{\lambda}^\prime$, there is some freedom in
the choice of the initial and final points. Whenever convenient, this
freedom will be used to choose $\lambda$ and $\lambda^\prime$ so
that their $z$ parameters have modulus greater than or equal to 1.
The following necessary conditions of deterministic
transformations can be expressed in a simple way when such a
choice is made.\\
\emph{Corollary: Let $\lambda=(z;c_1,\ldots,c_p)$ and
$\lambda^\prime=(z^\prime;c_1^\prime,\ldots,c_p^\prime)$ be truly
multipartite such that $\abs{z}\geq1$ and $\abs{z^\prime}\geq1$.
If $\lambda$ can be transformed into $\lambda^\prime$ by LOCC,
then $\abs{z^\prime}\geq\abs{z}$ and $c_k^\prime\geq c_k$ for all
$k$.}

This statement follows straightforwardly from theorem 2. The set
of sufficient conditions, however, are expressed differently
depending on whether the initial and the final states have a
vanishing cosine or not. A separate analysis has to be given in
each special case, which can be found in the following three
subsections.

\subsection{Transformations into states with a
vanishing cosine}

If $\lambda$ to $\lambda^\prime$ conversion is possible and
$\lambda^\prime$ has vanishing cosines, then $\lambda$ should also
have vanishing cosines for the same parties. Hence, this case
deals with transformations between states with vanishing cosines.
In this case, it turns out that the necessary conditions given
in the corollary above are also sufficient.\\
\emph{Theorem 3. Let $\lambda=(z;c_1,\ldots,c_p)$ and
$\lambda^\prime=(z^\prime;c_1^\prime,\ldots,c_p^\prime)$ be truly
multipartite, $\abs{z}\geq1$, $\abs{z^\prime}\geq 1$, and
$\lambda^\prime$ has a vanishing cosine parameter. Then $\lambda$
can be LOCC converted into $\lambda^\prime$ if and only if
$\abs{z^\prime}\geq\abs{z}$ and $c_k^\prime\geq c_k$ for all
parties $k$.}

\textit{Proof:} Necessity is obvious from the corollary. For
proving the sufficiency of the conditions, it must be shown that
any desired parameter (one of the cosines or the $z$ parameter)
can be increased without changing the others. In order to
simplify the proof, suppose that the first party has a
vanishing cosine in $\lambda^\prime$ and hence $c_1^\prime=c_1=0$.
Suppose also that both $z$ and $z^\prime$ are positive real numbers.

First, note that the first party can increase
the $z$ parameter without changing any of the cosines. In other
words, the state $\lambda=(z;0,c_2,\ldots,c_p)$ can be converted
to $(z^\prime;0,c_2,\ldots,c_p)$ for any real number $z^\prime$
with $z^\prime>z$. In terms of the first parametrization of local
measurements, this can be achieved by a two outcome measurement,
having the following parameters
\begin{eqnarray}
  A_1  &=& \sqrt{\frac{z^{\prime2}z^2-1}{z^{\prime4}-1}} \quad,\\
  A_2  &=& z^\prime \sqrt{\frac{z^{\prime2}-z^2}{z^{\prime4}-1}} \quad,\\
  B_1  &=& \frac{z^\prime}{z}\sqrt{\frac{z^{\prime2}z^2-1}{z^{\prime4}-1}} \quad,\\
  B_2  &=& \frac{1}{z} \sqrt{\frac{z^{\prime2}-z^2}{z^{\prime4}-1}} \quad,\\
  C_1 &=& C_2=\gamma_1=\gamma_2=0\quad.
\end{eqnarray}
It can be easily seen that the conditions
(\ref{eq:Par1}-\ref{eq:Par3}) are satisfied by these parameters
and both of the final points are LU-equivalent to
$(z^\prime;0,c_2,\ldots,c_p)$.

Second, any party other than the first can increase their own
cosine parameter to any desired value without changing any other
parameter. For this purpose, consider $k$th party with $k\neq1$
whose initial and final cosines satisfying $c_k^\prime > c_k\geq0$.
The $k$th party can carry out the following two-outcome measurement, which is expressed in the first
parametrization as,
\begin{eqnarray}
  A_1 &=& A_2 = B_1 = B_2 =\frac{1}{\sqrt{2}}\quad,\\
  C_1 &=& C_2 = c_k^\prime\quad,\\
  \gamma_1 &=& - \gamma_2 = \arccos\frac{c_k}{c_k^\prime}\quad.
\end{eqnarray}
It is straightforward to check that the conditions
(\ref{eq:Par1}-\ref{eq:Par3}) are satisfied. If the initial state point
is $(z;0,c_2,\ldots,c_{k-1},c_k,c_{k+1},\ldots,c_p)$, then the final
point for both outcomes is LU-equivalent to
$(z;0,c_2,\ldots,c_{k-1},c_k^\prime,c_{k+1},\ldots,c_p)$.

For transforming $\lambda$ into $\lambda^\prime$, the first party
increases only the modulus of the $z$ parameter while the rest
of the parties increase their cosines. Note that all of these
local operations are deterministic. Moreover, they can be carried
out in any order without changing the operation parameters. $\Box$

\subsection{Transformations from states with non-zero cosines}

Next case that will be dealt with is transformations between
states without vanishing cosines. For such transformations, the
following representation of the complex $z$-parameters of the
points of $\Lambda$ turns out to be very useful. As a result, a
brief explanation of that representation is necessary at this
point. Let $z$ be a complex number having the polar decomposition
$z=\exp(\rho+i\theta)$. Two real valued functions $n=n(z)$ and
$s=s(z)$ are defined as
\begin{eqnarray}
  n &=& \frac{\cos\theta}{\cosh\rho}=\frac{2\,\mathrm{Re}\,z}{\abs{z}^2+1}\quad,\\
  s &=& \frac{\sin\theta}{\sinh\rho}=\frac{2\,\mathrm{Im}\,z}{\abs{z}^2-1}\quad.
\end{eqnarray}
Note that $n$ takes on values in the closed interval
$[-1,1]$ while $s$ takes on values in the closed interval
$[-\infty,+\infty]$. In particular, $s$ has the value $\pm\infty$
on the unit circle $\abs{z}=1$. At the special points $z=\pm1$ of
the unit circle, however, $s$ does not have a definite value or
limit. Fortunately, these two points are the only places where $n$
reaches its boundary values, namely $n=+1$ only at $z=1$.
Similarly, $n=-1$ only at $z=-1$.

The correspondence between $z$ and the pair $(n,s)$ is two-to-one
for all points on the complex plane except $z=\pm1$. First note
that, if $z$ is replaced with $1/z$, then both of these two
functions do not change: $n(1/z)=n(z)$ and $s(1/z)=s(z)$. The
opposite is also true, i.e., if $n(z)=n(z^\prime)$ and
$s(z)=s(z^\prime)$ then we either have $z=z^\prime$ or
$z=1/z^\prime$. Hence, the pairs of values $(n,s)$ are identical
for conjugate points of $\Lambda$.

Now, consider the transformation of $\lambda$ into
$\lambda^\prime$ where all of the cosines of $\lambda$ are
non-zero. In that case, all cosines of $\lambda^\prime$ should be
non-zero as well if a transformation is possible. Since both
states are considered to be truly multipartite, both of them have
at most two LU-equivalent points in $\Lambda$. The following
theorem gives the necessary and sufficient conditions for the
LOCC transformation between such states.\\
\emph{Theorem 4. Let $\lambda=(z;c_1,\ldots,c_p)$ and
$\lambda^\prime=(z^\prime;c_1^\prime,\ldots,c_p^\prime)$
correspond to truly multipartite states and all cosines of both
states are non-zero. Let $(n,s)$ and $(n^\prime,s^\prime)$ denote
the values of the $n$ and $s$ functions of the $z$ parameters of
$\lambda$ and $\lambda^\prime$ respectively. It is possible to
transform $\lambda$ into $\lambda^\prime$ by LOCC if and only if
\begin{itemize}
\item[(a)] $c_k^\prime\geq c_k$ for all parties $k$, and
\item[(b)] the following equality is satisfied
  \begin{equation}
    \frac{n^\prime}{n} = \frac{s^\prime}{s} = \frac{c_1c_2\ldots c_p}{c_1^\prime c_2^\prime\ldots c_p^\prime}\quad.
  \end{equation}
\end{itemize}}

\textit{Proof:} First we show necessity. If $\lambda$ can be
converted into $\lambda^\prime$, then part (a) follows from the
corollary. The relation in (b) can be obtained from the extension
of the relations (\ref{eq:Rel1}-\ref{eq:Rel3}) into the whole
protocol as follows: Let $L$ denote the chain of outcomes of all
local operations carried out by all parties until the protocol is
terminated and let $p_L$ denote the joint probability of
occurrence of that outcome. After carefully following the second
parametrization of all successive local operations, one reaches
to a final parameter point
$\lambda^{(L)}=(z^{(L)};c_1^{(L)},\ldots,c_p^{(L)})$ at the end of
the protocol for the outcome $L$. Let $c(\lambda^{(L)})$ denote
the product of all cosines of this state, i.e.,
$c(\lambda^{(L)})=c_1^{(L)}\cdots c_p^{(L)}$. Now, the successive
use of relations (\ref{eq:Rel1}-\ref{eq:Rel3}) immediately lead to
\begin{eqnarray}
  \sum_L p_L  \frac{1}{N(\lambda^{(L)})} &=&  \frac{1}{N(\lambda)}\quad,    \label{eq:AdvRel1} \\
  \sum_L p_L  \frac{\abs{z^{(L)}}^2}{N(\lambda^{(L)})} &=&  \frac{\abs{z}^2}{N(\lambda)} \label{eq:AdvRel2}\quad, \\
  \sum_L p_L  \frac{z^{(L)}c(\lambda^{(L)})}{N(\lambda^{(L)})} &=& \frac{z c(\lambda)}{N(\lambda)} \label{eq:AdvRel3}\quad.
\end{eqnarray}
All of these relations are valid for all probabilistic
transformations as well. However, for the current deterministic
transformation, all final states can be either
$\lambda^{(L)}=\lambda^\prime$ or
$\lambda^{(L)}=\hat{\lambda^\prime}$. Hence, all terms within the
summation can be collected into just two terms with total
probabilities $p$ and $q=(1-p)$ respectively. Using,
$N(\hat{\lambda^\prime})=N(\lambda^\prime)/\abs{z^\prime}^2$, the
relations above can be expressed as
\begin{eqnarray}
   \frac{p+q\abs{z^\prime}^2}{N(\lambda^\prime)} &=&  \frac{1}{N(\lambda)}\quad, \\
  \frac{p\abs{z^\prime}^2+q}{N(\lambda^\prime)}  &=&  \frac{\abs{z}^2}{N(\lambda)}\quad, \\
  \frac{p z^\prime +q z^{\prime*}}{N(\lambda^\prime)} c(\lambda^\prime) &=& \frac{z c(\lambda)}{N(\lambda)}\quad.
\end{eqnarray}
(Note that these equations are valid for the cases $z^\prime=\pm1$
as well, for which $\hat{\lambda^\prime}=\lambda^\prime$ and there
should only be a single term. For these special cases, the
equation above holds for all possible probabilities $p$.) These
equations are equivalent with the following four equations
\begin{eqnarray}
  \frac{N(\lambda^\prime)}{N(\lambda)} &=& \frac{\abs{z^\prime}^2+1}{\abs{z}^2+1}\quad, \label{eq:Result1} \\
  (p-q)\frac{\abs{z^\prime}^2-1}{\abs{z^\prime}^2+1} &=& \frac{\abs{z}^2-1}{\abs{z}^2+1}\quad,\\
  \frac{\mathrm{Re}\,z^\prime}{\abs{z^\prime}^2+1} c(\lambda^\prime) &=&  \frac{\mathrm{Re}\,z}{\abs{z}^2+1} c(\lambda) \quad,\label{eq:Result3}\\
  (p-q)\frac{\mathrm{Im}\,z^\prime}{\abs{z^\prime}^2+1} c(\lambda^\prime) &=&  \frac{\mathrm{Im}\,z}{\abs{z}^2+1} c(\lambda) \quad. \label{eq:Result4}
\end{eqnarray}
Expressing the last three relations in terms of $n$ and $s$, we
get the desired relation. This completes the proof of necessity.
(Again, note that for the special cases $z^\prime=\pm1$, the
equations do not depend on the precise value of $p-q$.)

\begin{figure}
\includegraphics[scale=0.8]{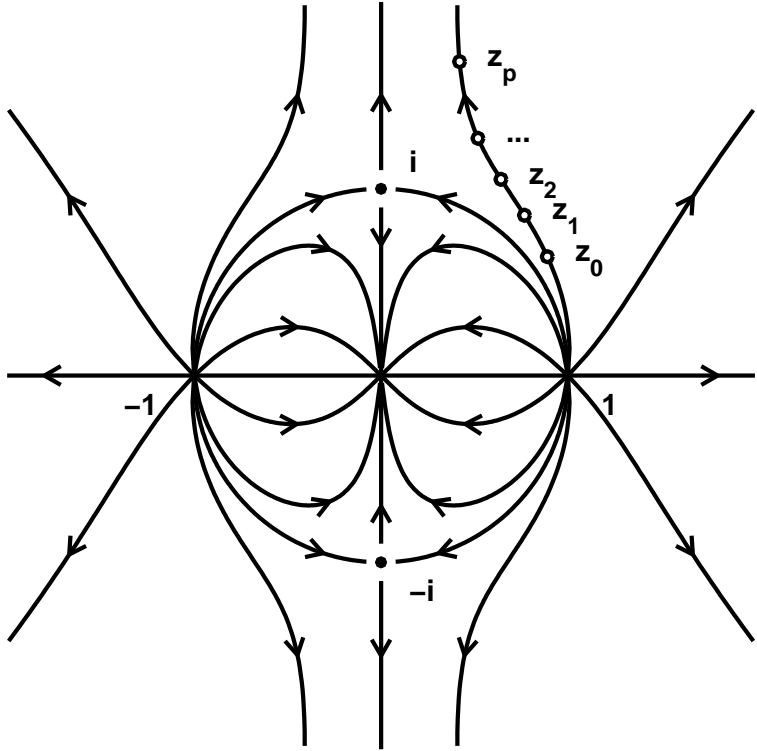}
\caption{The curves in the complex plane showing the $z$-parameter
values that are left invariant by deterministic LOCC
transformations (i.e., curves for which $n/s=$const.) and the
direction of the shift of the $z$-parameter under these
transformations. The successive $z$-parameter values for the
conversion protocol are also shown. Here, the local
operation of party-$k$ changes that parameter from $z_{k-1}$ to
$z_k$.} \label{fig:z}
\end{figure}

For proving sufficiency of the conditions, it will be argued that
the parties consecutively make a deterministic transformation by a
local operation to bring the initial state to some desired final
state. Hence, there will be a chain of points $\lambda_k$ for
$k=0,1,\ldots,p$,
$$ \lambda_0=\lambda \rightarrow
\lambda_1\rightarrow\lambda_2\rightarrow \cdots\rightarrow
\lambda_p=\lambda^\prime
$$
where party-$k$ takes the $k$th turn to change the state point
from $\lambda_{k-1}$ into $\lambda_{k}$. Here, the intermediate
points are given as
\begin{eqnarray}
  \lambda_0 &=& (z_0;c_1,c_2,c_3,\ldots,c_p)\quad,\nonumber\\
  \lambda_1 &=& (z_1;c_1^\prime,c_2,c_3,\ldots,c_p)\quad,\nonumber\\
  \lambda_2 &=& (z_2;c_1^\prime,c_2^\prime,c_3,\ldots,c_p)\quad,\nonumber\\
  \cdots & & \cdots \nonumber\\
  \lambda_p &=& (z_p;c_1^\prime,c_2^\prime,c_3^\prime,\ldots,c_p^\prime)\quad,\nonumber
\end{eqnarray}
where $z_0=z$ and $z_p=z^\prime$. The $k$th party essentially
increases her cosine from $c_k$ to $c_k^\prime$ while this change
is associated by a definite change in the value of the
$z$-parameter from $z_{k-1}$ to $z_{k}$ (see Fig.~\ref{fig:z}).
The intermediate values $z_k$ of that parameter should be found
from the condition (b) of the theorem, e.g.,
$n(z_k)=n(z_{k-1})c_k/c_k^\prime$ etc. Hence, what is left is the
proof that, for all $k$, the transformation from $\lambda_{k-1}$
to $\lambda_{k}$ by the $k$th party can be carried out. Obviously,
only the cases for which $c_k^\prime>c_k$ are needed to be
considered.


Unfortunately there are special values of the $z$-parameter that
need a separate treatment. Note that, due to the proved necessity
of the condition (b), the special values of $0$ and $\infty$ for
$n$ or $s$ cannot change in these deterministic transformations.
Specifically, these correspond to (a) the real axis,
$\mathrm{Im}\,z=0$, where $s=0$; (b) the imaginary axis,
$\mathrm{Re}\,z=0$, where $n=0$; (c) and the unit circle,
$\abs{z}=1$ where $s=\pm\infty$. These are curves that are
invariant under deterministic LOCC transformations. Hence, their
intersections, specifically $z=\pm1,\pm i$ need special attention.

Now, consider the local operation done by the $k$th party to
transform the state point from $\lambda_{k-1}$ to $\lambda_{k}$. A
deterministic transformation with two outcomes $\pm$ will do the
job, which is designed, depending on the special cases, as follows.
\begin{itemize}
\item[Case (I)] \emph{When $z_{k-1}$ and $z_k$ are not on the unit circle,
i.e., with $z_m=\exp(\rho_m+i\theta_m)$, both $\rho_{k-1}$ and
$\rho_k$ are strictly positive:} In the second parametrization of
local operations, the parameters of transformation are
\begin{eqnarray}
  p_{\pm} &=& \frac{1}{2}\left(1\pm\frac{\tanh\rho_{k-1}}{\tanh\rho_{k}}\right) \quad,\\
  C_{\pm} &=& c_k^\prime \quad,\\
  z^{(\pm)} &=& (z_k)^{\pm1}\quad.
\end{eqnarray}
Now, it is straightforward, but tedious, to check that $p_\pm$ are
probabilities and the parameters given above satisfy the
conditions (\ref{eq:Rel1}) and (\ref{eq:Rel3}). Finally, it is
trivial to see that the final state is LU-equivalent to
$\lambda_{k}$.

\item[Case (II)] \emph{Either $z_{k-1}$ or $z_k$ are on the unit circle (in
which case both points must be on the unit circle and therefore
$\rho_{k-1}=\rho_{k}=0$), but both points are different from
$\pm1,\pm i$ (i.e., $\theta_{k-1}$ and $\theta_{k}$ are
not an integer multiple of $\pi/2$):} In this case, the
parametrization above in Case (I) is valid, except that the
probabilities should now be expressed in terms of the polar angles as
\begin{equation}
 p_{\pm} = \frac{1}{2}\left(1\pm\frac{\tan\theta_{k-1}}{\tan\theta_{k}}\right)\quad.
\end{equation}
Here too, it is straightforward to check that this local operation
describes the needed transformation.

\item[Case (III)] \emph{Either $z_{k-1}=\pm i$ or $z_{k}=\pm
i$:} Since the complex numbers $\pm i$ have $(n,s)$ parametrization
given by $(n,s)=(0,\pm\infty)$, by part (b) of the current
theorem, this point cannot be changed by deterministic LOCC
transformations. Hence both of the $z$ parameters should be $\pm
i$. Moreover, as $z$ parameters of points in $\Lambda$, $i$ and
$-i$ correspond to LU-equivalent points. Hence, take
$z_{k-1}=z_{k}=i$ without loss of generality. The main idea is
that, even though the $z$ parameter does not change, the $k$th
party can increase her cosine for this special case. The
parameters of the transformation are given as
\begin{eqnarray}
  p_\pm &=& \frac{1}{2}\left(1\pm\frac{c_k}{c_k^\prime}\right)\quad,\\
  C_\pm &=& c_k^\prime\quad,\\
  z^{(\pm)} &=& \pm i\quad.
\end{eqnarray}

\item[Case (IV)] \emph{When $z_{k}=\pm 1$:} That final point has an
extreme $n$ value of $\pm1$. Hence, by part (b) of the current theorem,
the only way this final point is reached is that $z_{k-1}=z_{k}$
and $c^\prime_k=c_k$. In other words $\lambda_{k-1}=\lambda_{k}$
and there is no need for a transformation.

\item[Case (V)] \emph{When $z_{k-1}=\pm 1$:} Note that by part (b)
of the current theorem, the point $z_{k}$ must satisfy
$n(z_{k})=\pm(c_k/c_k^\prime)$. However, due to the fact that
$s(z_{k-1})$ does not have a definite value, $s(z_{k})$ is
arbitrary. Hence, suppose that $z_{k}$ is any number on the
complex plane such that
\begin{equation}
  n(z_{k})=\frac{2\mathrm{Re}\,z_{k}}{\abs{z_{k}}^2+1} =  \pm\frac{c_k}{c_k^\prime} \quad.
\end{equation}
Then, the parameters of the local measurement by the $k$th party in second
parametrization are given by
\begin{eqnarray}
  p_\pm &=& \frac{1}{2} \quad,\\
  C_\pm &=& c_k^\prime\quad,\\
  z^{(\pm)} &=& (z_k)^{\pm1}\quad.
\end{eqnarray}
Once it is observed that
\begin{eqnarray}
 N(\lambda_{k}) &=& 1+\abs{z_{k}}^2+c_1^\prime\cdots c_k^\prime c_{k+1}\cdots c_p (2\mathrm{Re}\,z_{k})\quad,\\
        &=& \left(1+\abs{z_{k}}^2\right)
            \left(1+c_1^\prime\cdots c_{k-1}^\prime c_k c_{k+1}\cdots c_p \,n(z_{k})\frac{c_k^\prime}{c_k}\right)\quad,\\
        &=& \left(1+\abs{z_{k}}^2\right)\frac{N(\lambda_{k-1})}{2}
\end{eqnarray}
it becomes straightforward to verify that the relations
(\ref{eq:Rel1}) and (\ref{eq:Rel3}) are satisfied and the desired
final state is produced. $\Box$

\end{itemize}

\subsection{Transformations from states with vanishing cosines to
those without any vanishing cosine}

The only remaining special case that is not yet dealt with is the
transformation of states with vanishing cosines to states having
only non-zero cosines. For these transformations to be possible,
it appears that the $z$-parameters of both the initial and the
final states must satisfy the following, state-independent restrictions.
\\
\emph{Theorem 5. Let $\lambda=(z;c_1,\ldots,c_p)$ and
$\lambda^\prime=(z^\prime;c_1^\prime,\ldots,c_p^\prime)$ be truly
multipartite states such that $\lambda$ has a vanishing cosine and
$\lambda^\prime$ has no vanishing cosines. It is possible to
transform $\lambda$ to $\lambda^\prime$ by LOCC if and only if
\begin{itemize}
\item[(a)] $c_k^\prime\geq c_k$ for all $k$,
\item[(b)] $\abs{z}=1$.
\item[(c)] $z^\prime$ is purely imaginary.
\end{itemize}
}

\textit{Proof:}  For proving the necessity of the conditions, the
relations (\ref{eq:Result1}-\ref{eq:Result4}) can be used.
Inserting $c(\lambda)=0$ and $c(\lambda^\prime)\neq0$ into the
last two equations we get $\mathrm{Re}\,z^\prime=0$ and $p=q$.
This then leads to $\abs{z}=1$. Part (a) follows from the
corollary again.

For proving sufficiency, first suppose that the first party has
vanishing cosine for the initial state, i.e., $c_1=0$. In theorem
3, it is shown that all parties except the first can increase
their cosines to any desired value without changing anything else.
Hence, the initial state $\lambda=(z;0,c_2,\ldots,c_p)$ can be
transformed into
$\tilde{\lambda}=(z;0,c_2^\prime,\ldots,c_p^\prime)$. At this
point, 1st party can do a single measurement and change the state
from $\tilde{\lambda}$ into $\lambda^\prime$. The needed
measurement has two outcomes with the following parameters in the
second parametrization
\begin{eqnarray}
  p_\pm &=& \frac{1}{2}\quad,\\
  C_\pm &=& c_1^\prime\quad,\\
  z^{(\pm)} &=& (z^\prime)^{\pm1}\quad.
\end{eqnarray}
It can be checked that the relations (\ref{eq:Rel1}) and
(\ref{eq:Rel3}) are satisfied and the measurement produces the
desired final state.$\Box$

\subsection{Invariants of deterministic LOCC transformations}

Having found all of the necessary and sufficient conditions for
deterministic LOCC transformations of truly multipartite states
into each other, a few remarks can be made about some interesting
features of these transformations. The most important of these is
the existence of some invariants. Many of these can be derived
from the particular relation in condition (b) of theorem 4. For
example, the phase angle of $(z-z^{-1})$ modulo $\pi$ is an
invariant for states having only non-zero cosines.

Although the deterministic transitions from a given state
$\lambda$ are allowed only to a restricted set of states, it might
be useful to consider also the sets of states that can be
transformed into $\lambda$. With this in mind, it can be seen that
the set of truly multipartite states can be partitioned into
various disjoint sets, between which no deterministic
transformations are possible. An invariant that is capable of
finely describing such a partition is given by
\begin{eqnarray}
  \xi(\lambda)&=& c_1c_2\cdots c_p~\frac{z+z^*}{1+\abs{z}^2} \\
     &=& c_1\cdots c_p n(z) = \frac{N(\lambda)}{1+\abs{z}^2}-1\quad,
\end{eqnarray}
for $\lambda=(z;c_1,\ldots,c_p)$. The existence of such an
invariant has been first discussed by Spedalieri\cite{Spedalieri}.
The invariance of this quantity follows directly from
Eq.~(\ref{eq:Result1}) which is valid for all deterministic
transformations into truly multipartite states. Note that
$\xi(\lambda)$ has the same value for all LU-equivalent truly
multipartite points of $\Lambda$. Note also that $\xi(\lambda)$
takes on values in the open interval $(-1,1)$.

For any $\xi$ with $-1<\xi<1$, define $M_\xi$ to be the set of all
truly multipartite $\lambda$ for which $\xi(\lambda)=\xi$. Hence,
any element of $M_\xi$ can only be deterministically converted
\emph{into} or \emph{from} some element of the same set. This is
the finest partition of the rank 2 states having that property.
Any state point $\lambda$ (or its whole LU-equivalence class) in
these sets will called as an \emph{ancestor} if it cannot be
obtained deterministically from a different state. In that
respect, the ancestors can be thought as \emph{the most entangled}
states, meaning that there are no other candidates for being even
more entangled. All states represented in $M_\xi$ can be obtained
from one of the ancestors but it appears that some of these sets
contain different LU-inequivalent ancestors.

The transformations within the set $M_0$ are treated in all three
of the theorems 3, 4 and 5. This set contains the states that have
either a vanishing cosine or a purely imaginary $z$ parameter. All
states in this set can be obtained from a single ancestor, the
GHZ state $\lambda_{\mathrm{GHZ}}=(1;0,0,\ldots,0)$, in
other words states which are LU-equivalent to Eq.~(\ref{eq:GHZ}).
An interesting special subset of $M_0$ is formed from states with
non-zero cosines having a $z$-parameter equal to $\pm i$, i.e.,
the set $L_i$, which is defined as
\begin{equation}
  L_i=\{(z;c_1,c_2,\ldots,c_p): c_1\cdots c_p\neq 0 ~,~z=\pm i\}
\end{equation}
Note that $L_i$ is also invariant under deterministic LOCC. If a
state in $L_i$ is transformed deterministically to a truly
multipartite state, then the final state must be in $L_i$ as well.
As a result, only the cosines of the state can be increased in a
deterministic transformation. It is not possible to change the $z$
parameter to any value other than $\pm i$.

The transformations within the sets $M_\xi$ for which $\xi\neq0$ are
covered in theorem 4. These sets contain infinitely many,
LU-inequivalent ancestors. The ancestors are those members of
$M_\xi$ that have a $z$ parameter equal to $+1$ (if $\xi>0$) or
$-1$ (if $\xi<0$). Note that a given ancestor can generate through
deterministic LOCC transformations only a subset of the points in
$M_\xi$. Moreover, all non-ancestor states can be generated from
different ancestors. Hence, the partial order in $M_\xi$ induced
by the LOCC-convertibility relationship is very non-trivial.

\section{Conclusions}
\label{sec:conc}

All of the necessary and sufficient conditions for the possibility
of converting truly-multipartite rank-2 states into each other are
given. The main theorems are listed under three different headings
depending on whether the states have vanishing cosines or not. It
is found that the multipartite states can be partitioned into
disjoint subsets, which are defined by a single continuous
parameter, in such a way that all deterministic transformations
are allowed within each subset only. The ancestor states, which
can be considered as the most entangled states for deterministic
transformations, are identified. They are either the GHZ state or
the states that correspond to $\lambda$ for which the
$z$-parameter is $\pm1$ and all cosines are non-zero.

For allowed transformations, a specific protocol for converting
the initial state to the final one is also proposed. In all of the
special cases investigated, the whole transformation can be
divided into a series of $p$ successive steps. Each step is
associated with one of the parties and is a deterministic
transformation by itself. In each step, the associated party
carries out a local generalized measurement; subsequently informs
all other parties about the outcome by classical communication;
and finally, all parties apply an appropriate local unitary
transformation. Depending on the initial and final states, some of
the parties may not need to do carry out any measurement in their
steps but every party makes \emph{at most one} such measurement.
Moreover, it can be shown that the precise ordering of the parties
can be changed arbitrarily; but the individual steps of local
operations may depend on the ordering.

These protocols have sufficient simplicity so that a comparison to
transformations of bipartite states can be made. Any deterministic
or probabilistic transformation between bipartite states can be
carried out with a protocol where (i) one of the parties carries
out a single general measurement, (ii) sends the outcome to the
other party by one-way classical communication, (iii) and the
other party subsequently carries out a local unitary
transformation\cite{lo_popescu}. It does not matter which party,
the 1st or the 2nd, does the local measurement.

For the transformations of truly multipartite states, it is
obvious that, in general all parties must carry out a non-trivial
local measurement. For example, if the $k$th cosines of initial
and final states are different, then party-$k$ must carry out a
local measurement because there is no other way for changing that
parameter. Associated with this necessity, each party must be able
to send classical information to all of the other parties for the
generic case, i.e., multi-way classical communication is also
needed. Hence, keeping these restrictions in mind, it can be
argued that the protocols proposed in this article, which require
\emph{at most one local operation for each party}, are the
simplest possible protocols. This feature is also present in the
protocol used in Ref. \onlinecite{GHZ_Distill} for the
distillation of the three-partite GHZ state with maximum
probability. An interesting question is this: for which kind of
multipartite entanglement transformations it is sufficient to
carry out the conversion by at most one local operation for each
party?

There are still unsolved problems in the LOCC transformation of
rank-2 states. One of those is the transformation into bipartite
entangled states and the other is the probabilistic
transformations. Hopefully, the approach taken in this article,
i.e., particular $\Lambda$ representation of the states and
the associated description of local operations will prove
useful in the solution of these problems as well.

\section*{Acknowledgement}

The authors are grateful to an anonymous referee for pointing out the work in
Ref.~\onlinecite{Spedalieri}.


\begin{thebibliography}{99}

\bibitem{Bennett_Telep} C. H. Bennett, G. Brassard, C. Cr\'{e}peau,
  R. Jozsa, A. Peres, and W. K. Wootters,
  Phys. Rev. Lett. \textbf{70}, 1895 (1993).

\bibitem{Bennett_Densecode} C. H. Bennett and S. J. Wiesner,
  Phys. Rev. Lett. \textbf{69}, 2881 (1992).

\bibitem{Bennett_Purif} C. H. Bennett, G. Brassard, S. Popescu,
  B. Schumacher, J. A. Smolin, and W. K. Wootters,
  Phys. Rev. Lett. \textbf{76}, 722 (1996).

\bibitem{Bennett_Conc} C. H. Bennett, H. J. Bernstein, S. Popescu,
   and B. Schumacher, Phys. Rev. A \textbf{53}, 2046 (1996).

\bibitem{lo_popescu} H.-K. Lo and S. Popescu,
   Phys. Rev. A \textbf{63}, 022301 (2001).

\bibitem{Nielsen_Maj} M. A. Nielsen, Phys. Rev. Lett. \textbf{83}, 436 (1999).

\bibitem{Vidal_Prob} G. Vidal, Phys. Rev. Lett. \textbf{83}, 1046(1999).

\bibitem{Plenio_Prob} D. Jonathan and M. B. Plenio, Phys. Rev. Lett. \textbf{83}, 1455 (1999).

\bibitem{Peres_Schmidt} A. Peres, Phys. Lett. A \textbf{202}, 16 (1995).

\bibitem{Bennett_Multi} C. H. Bennett, S. Popescu, D. Rohrlich, J. A. Smolin, and A. V. Thapliyal,
  Phys. Rev. A \textbf{63}, 012307 (2000).

\bibitem{Xin_Duan} Y. Xin and R. Duan, Phys. Rev. A \textbf{76}, 044301 (2007).

\bibitem{GHZ_Distill} A. Ac\'{i}n, E. Jan\'{e}, W. D\"ur, and G. Vidal,
  Phys. Rev. Lett. \textbf{85}, 4811 (2000).


\bibitem{Spedalieri} F. M. Spedalieri, arXiv:quant-ph/0110179 (2001).


\bibitem{Duan_tensor} E. Chitambar, R. Duan, and Y. Shi, Phys. Rev. Lett. \textbf{101}, 140502 (2008).

\bibitem{Eisert_Briegel} J. Eisert and H. J. Briegel, Phys. Rev. A \textbf{64}, 022306 (2001).

\bibitem{Wootters_Concur} W. K. Wootters, Phys. Rev. Lett. \textbf{80}, 2245 (1998).

\end{thebibliography}
\end{document}